Investigation of Convective Downburst Hazards to Marine Transportation
Mason, Derek
Thomas Jefferson High School for Science and Technology
Alexandria, VA

**Abstract**

Convective downbursts are known to produce potentially hazardous weather conditions.  Currently, severity indices are used to estimate the strength of a potential downburst, but this information does not readily translate to the variables affected by downburst events. The effects of downbursts are often associated with aviation because of rapid changes in wind direction and speed, but can also be observed in marine conditions.  Three recently observed downburst events have been selected as case studies to evaluate the effects of the downbursts on the marine environment.  The information gathered on these events includes wind speed, gusts and direction at the surface, air temperature and pressure, water level, and Wet Microburst Severity Index (WMSI) values.  The events selected occurred at Tolchester Beach, MD, on September 28, 2006 at 23:06 GMT, North Jetty, near Galveston, TX, on October 13, 2006 at 2:48 GMT, and again at North Jetty on October 27, 2006 at 5:30 GMT.  Correlation between the WMSI values, the maximum wind gust, and the change in water level is suggested.

**Introduction**

A downburst is a powerful downdraft produced by a convective storm that causes damaging winds for five to thirty minutes.  (Fujita, 1985) Once identified, downbursts can be divided into two categories, microburst and macroburst.  Winds from a microburst are horizontally limited (less than four kilometers), meaning that a smaller area is affected but also that higher wind speeds may result.  Winds from a macroburst spread farther than four kilometers but are more limited in wind speed. (Fujita, 1985) Downbursts are also divided into categories based on their execution such as traveling or stationary, wet or dry, or radial or twisting. Wet microbursts are distinguished from dry ones by the presence of heavy rain between the onset and offset of high wind gusts. (Wakimoto, 1985)

The effects of downbursts are often associated with aviation, but can also be observed in marine conditions. These conditions may include wind speed and direction at the surface, waves generated and their height, or water level changes. A downburst is capable of temporarily pushing these factors to their extremes, causing a potential hazard to property or human lives. Small craft such as water taxis or recreational vessels are very susceptible to high wind speeds and other factors that are intensified by the wind. A downburst is predictable for only a short time span and greatly enhances winds already being produced by a convective storm. This means that within a moderate storm a craft can be surprised by sudden high wind speeds.

Schwab (1978) concluded that changes in water level are affected heavily by wind speeds and that using wind speed forecast data it is possible to make predictions of storm surges with significant accuracy. Although pressure gradients are also associated with changes in water level, especially storm surge, wind speed has a greater effect than do pressure gradients when creating a storm surge in shallow areas, such as the port areas investigated in this paper. Although water level data is modified to ignore the normally insignificant oscillations created by waves, convective winds will bring waves to more hazardous heights as the water level increases. A downburst is not able to produce winds for a long enough period to cause a major storm surge, but the presence of a significant water level increase suggests that wave height and severity also increase.

The WMSI was developed by Pryor and Ellrod (2004) to predict the maximum wind gusts of a microburst resulting from convective activity. The index is calculated using the Theta-e deficit (TeD), representing downburst formation, and the convective available potential energy (CAPE), representing updraft strength. (Pryor and Ellrod, 2004) This index, when used to predict microburst strength, can be indirectly used to predict the subsequent rise in recorded water level after a microburst has occurred.

**Methodology**

Several resources were used to monitor downburst events and create a data set of the conditions. The Storm Prediction Center (SPC) displays high wind, tornado, and hail reports, and is used to locate areas of convective activity and high winds. Once an area of convective activity is identified, more specific observation programs are used. The Physical Oceanographic Real-Time System (PORTS) is a system of monitoring stations within certain port areas in the U.S. that records marine and some weather conditions. The National Data Buoy Center (NDBC) provides data similar to that from PORTS, but also shows the data from other observation systems in addition to its own. The data from these two sources is evaluated to determine whether or not a downburst did occur. It is also necessary to verify the presence of convective activity at the time of the downburst. This is done using NEXRAD data from the National Climatic Data Center at the time of the downburst event.

The WMSI is used to estimate microburst wind speeds should an event occur. Data values for the WMSI are recorded by the microburst product imagery provided by the National Environmental Satellite, Data, and Information Service (NESDIS) using Geostationary Operational Environmental Satellite (GOES) imagery and sounding data. WMSI images used in this study were generated with the magnification centered on the appropriate PORTS or NDBC station. The WMSI images are updated hourly and are not generated for some specific hours; therefore the closest WMSI values in time and location were recorded for this study.

**Case 1: Tolchester Beach, MD**

The data for this event was recorded by a PORTS station on September 28, 2006. The highest wind gust recorded was 49 knots at 23:06 GMT, followed by measurements of 43 and 35 knots at 23:12 and 23:18 GMT respectively. The water level measurement increased from 2.22 feet at 23:00 GMT to a peak of 3.16 feet at 23:24 GMT, and then fell back to 2.25 feet at 23:42 GMT. The air temperature falls from 73.94 $^{o}$F at 23:00 GMT to 61.34 $^{o}$F at 23:06 GMT, a 12.6 $^{o}$F decrease. The wind direction changes from 150$^{o}$ at 23:00 GMT to 276$^{o}$ at 23:06 GMT, a 126$^{o}$ difference. The air pressure increases from 1004.1 mb at 23:00 GMT to 1006.4 mb at 23:12 GMT, an increase of 2.3 mb. These results are graphed over a 3 day period automatically by PORTS (Fig. 1).

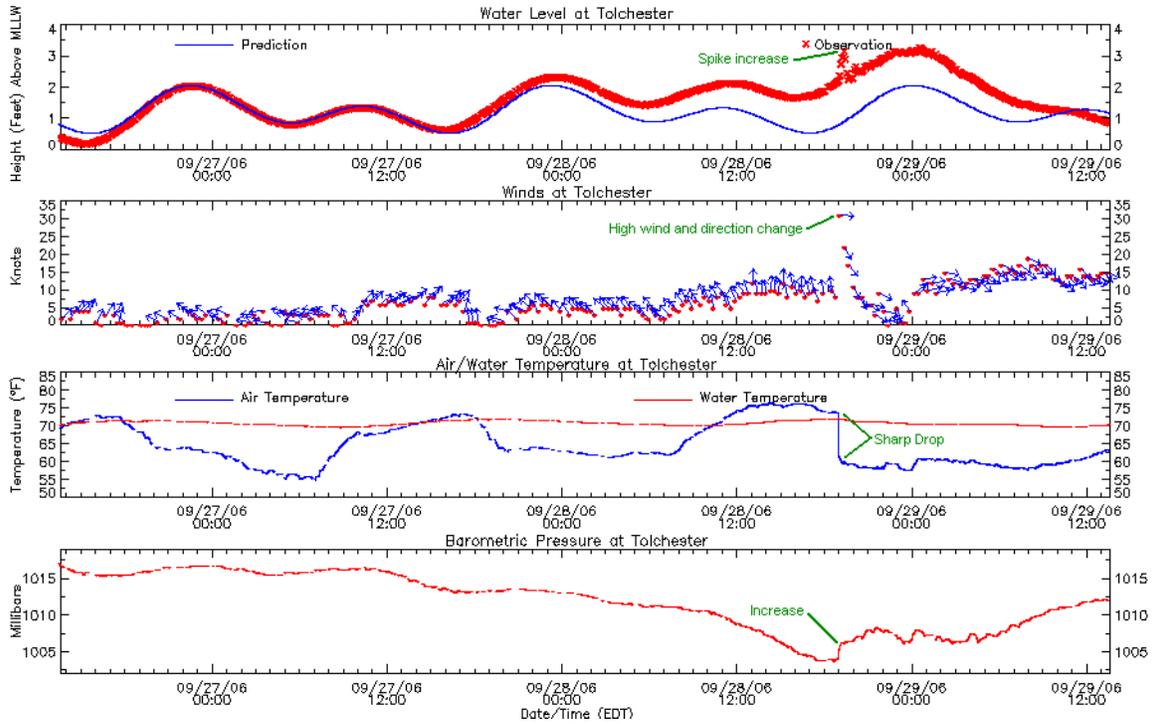

**Fig. 1** – 3-day meteorology from the PORTS station at Tolchester Beach, MD, from 9/27 to 9/29. Significant changes are identified in green.

The WMSI values were collected at 22:00 GMT. The closest value to the location of the downburst was 45 (Fig. 2). The downburst occurred while a large storm passed over the northern Chesapeake Bay region. A radar image of the area at 23:06 GMT verifies the convective activity (Fig. 3).

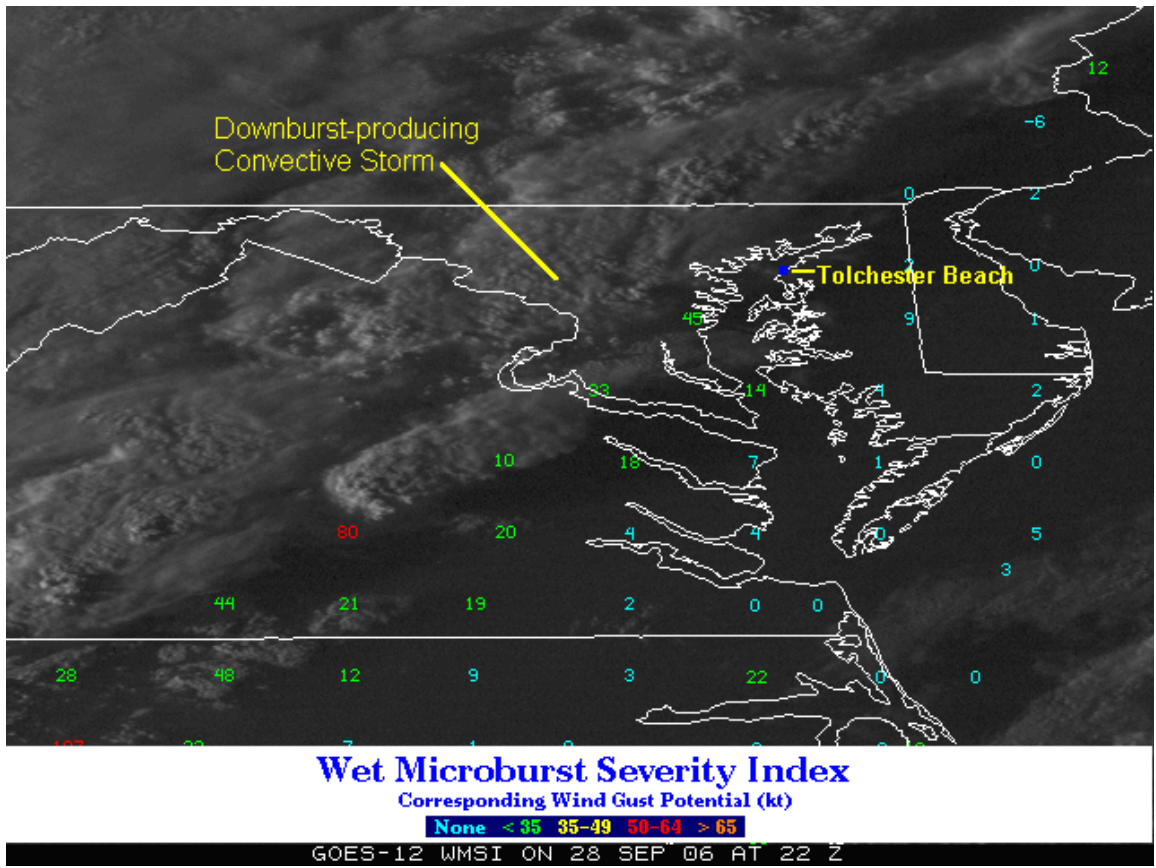

**Fig. 2** – WMSI values over the mid-Atlantic coast at 22:00 GMT. The Tolchester Beach station is located at the blue square.

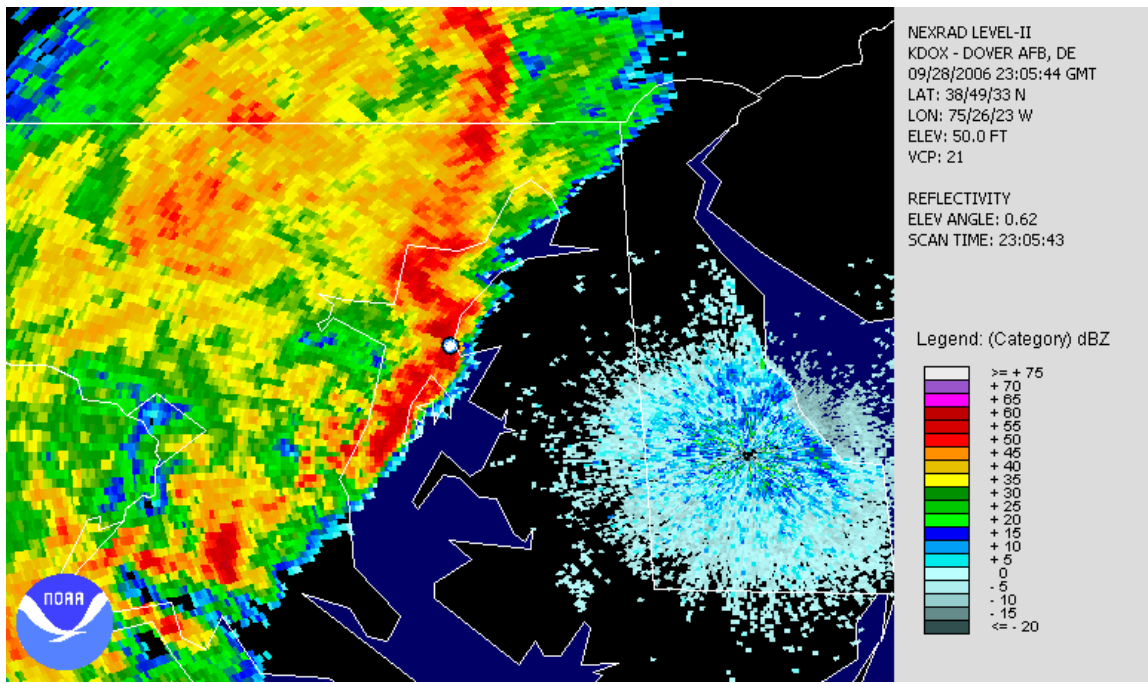

**Fig. 3** – Radar imagery from the NEXRAD station at Dover Air Force Base at 23:06 GMT. The Tolchester Beach station is located at the white dot.

**Case 2: North Jetty, TX**

The data for this event was collected from the PORTS station at North Jetty, TX, on October 13, 2006. A NDBC station also recorded supporting text data during this event, but part of the event is not recorded because data for 2:48 and 2:54 GMT is not recorded. The highest wind gust recorded was 35 knots at 2:54 GMT, which followed a measurement of 31 knots at 2:48 GMT. The water level measurement was under the predicted value since 13:18 GMT on October 10, 2006. However, it increased from 1.91 feet (0.08 below the predicted level) at 2:42 GMT to a peak of 2.28 feet (0.23 feet above the predicted level) at 3:06 GMT, and then fell back to 1.96 feet (0.17 feet below the predicted level) at 4:48 GMT. The air temperature falls from 78.44 $^{o}$F at 2:36 GMT to 69.98 $^{o}$F at 2:48 GMT, an 8.46 $^{o}$F decrease. The wind direction changes from 131$^{o}$ at 2:36 GMT to 2$^{o}$ at 2:42 GMT, a 133$^{o}$ difference. The pressure increased from 1009.4 mb at 2:30 GMT to 1010.1 mb at 2:42 GMT, a 0.7 mb increase. These results are graphed over a 3 day period automatically by PORTS (Fig. 4).

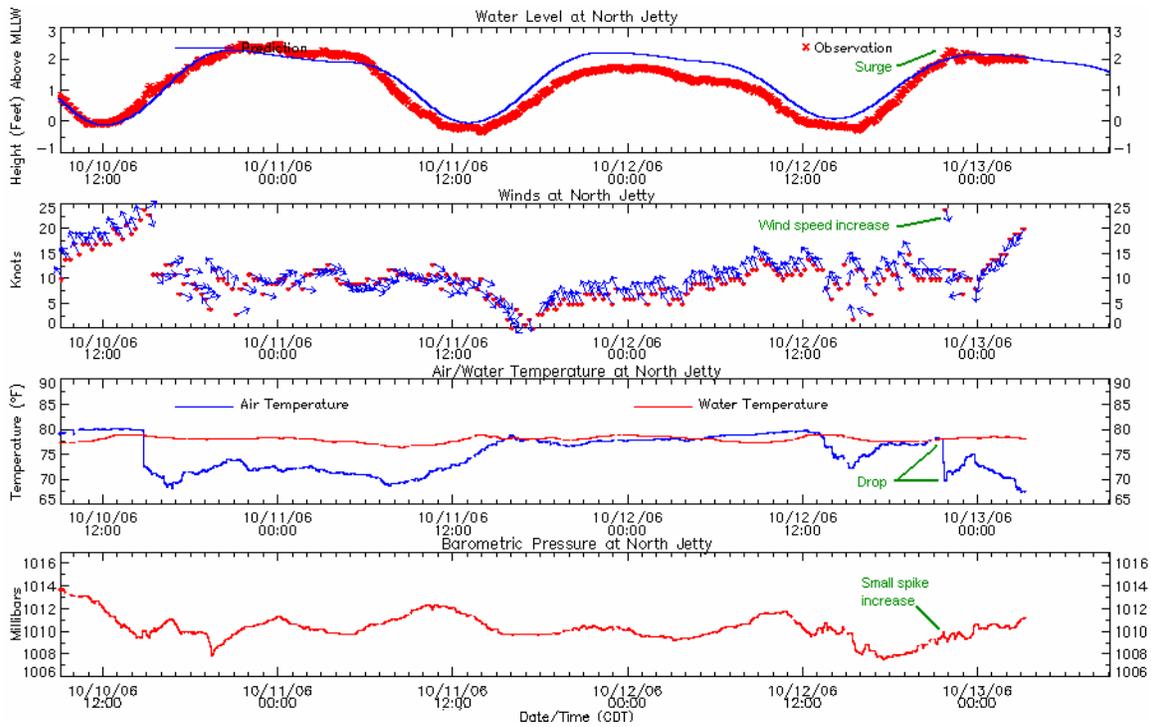

**Fig. 4** – 3-day meteorology from the PORTS station at North Jetty, TX, from 10/10 to 10/13. Significant changes are identified in green.

The WMSI values were collected at 2:00 GMT. The closest value to the location of the downburst was 77 (Fig. 5). The downburst occurred as a squall line passed near the station over the Gulf. A radar image of the area at 3:48 GMT verifies the convective activity (Fig. 6).

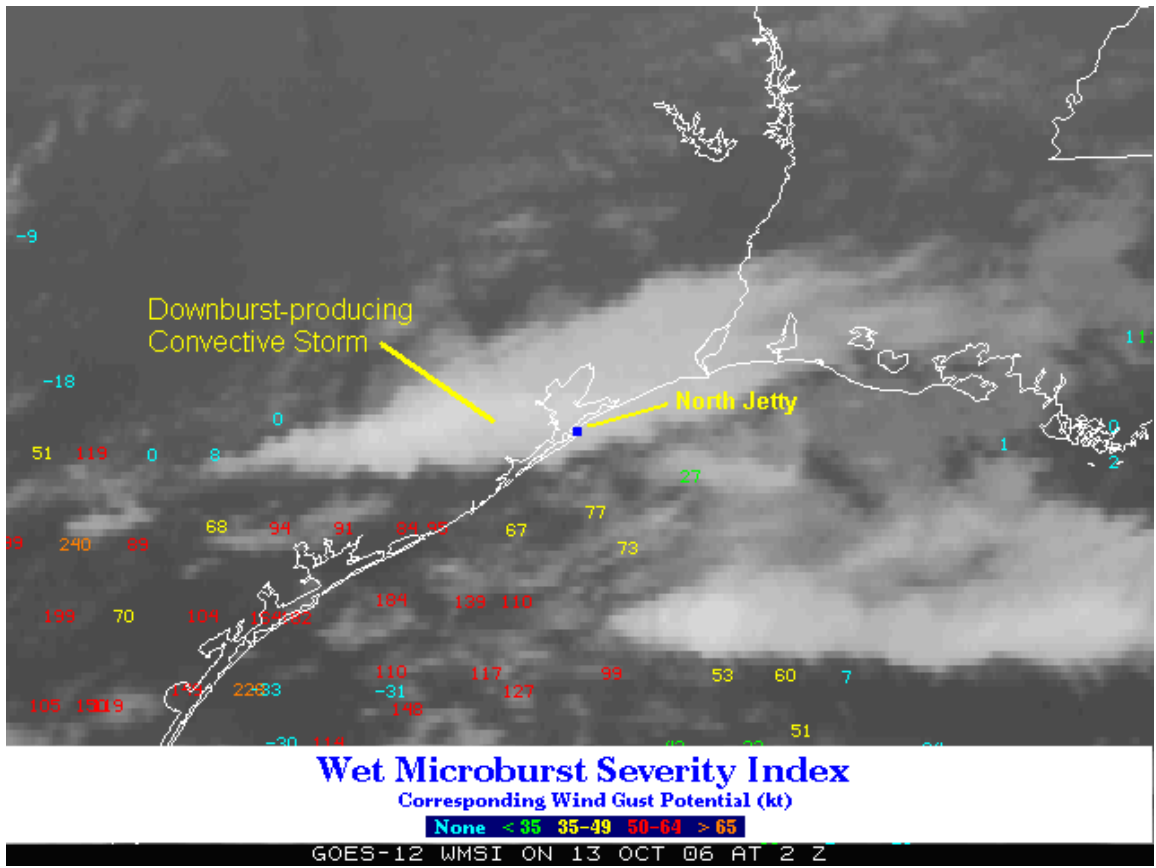

**Fig. 5** – WMSI values over the west Gulf coast at 2:00 GMT. The North Jetty station is located at the blue square.

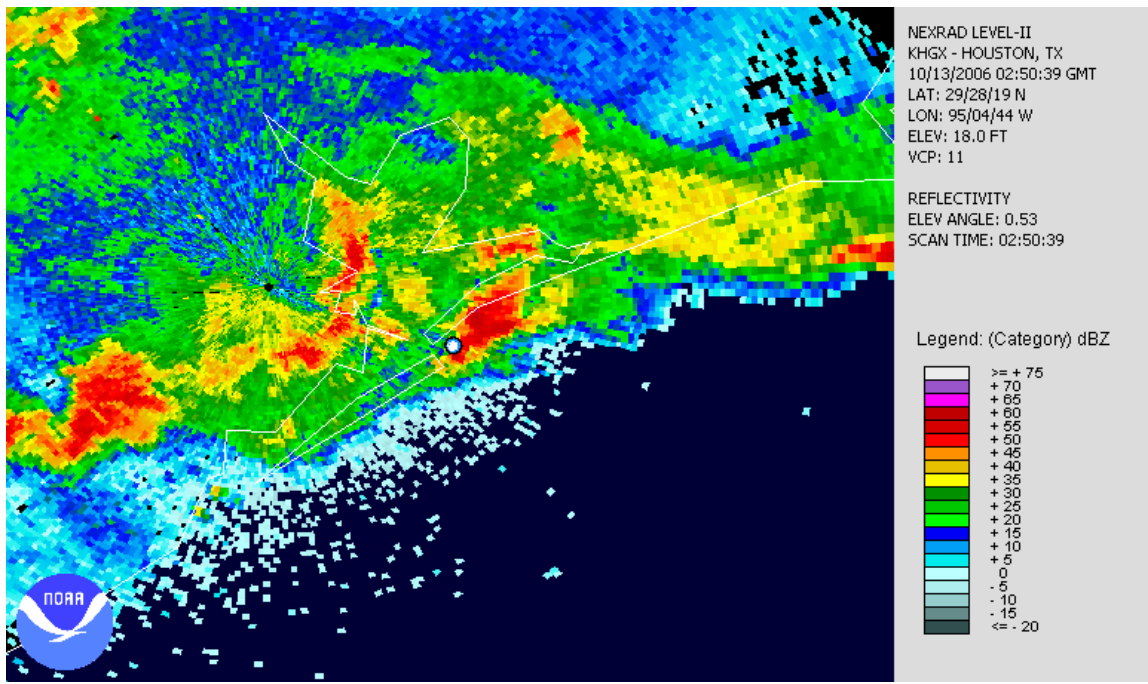

**Fig. 6** – Radar imagery from the NEXRAD station in Houston at 2:51 GMT. The North Jetty station is located at the white dot.

**Case 3: North Jetty, TX**

The data for this event was collected from the PORTS station at North Jetty, TX, on October 27, 2006. The downburst began with a recorded wind gust of 30 knots at 5:30 GMT and escalates to the highest wind gust recorded, 41 knots, at 5:36 GMT. The water level increased from 2.17 feet at 5:24 GMT to a peak of 2.31 feet at 5:36 GMT, and then fell back to 1.98 feet at 6:12 GMT. The air temperature falls from 76.10 °F at 5:24 GMT to 68.18 °F at 5:42 GMT, a 7.92 °F decrease. The wind direction changes from 179° at 5:24 GMT to 247° at 5:36 GMT, a 68° difference. The pressure increased from 1007.2 mb at 5:24 GMT to 1008.0 mb at 5:36 GMT, a 0.8 mb increase. These results are graphed over a 3 day period automatically by PORTS (Fig. 7).

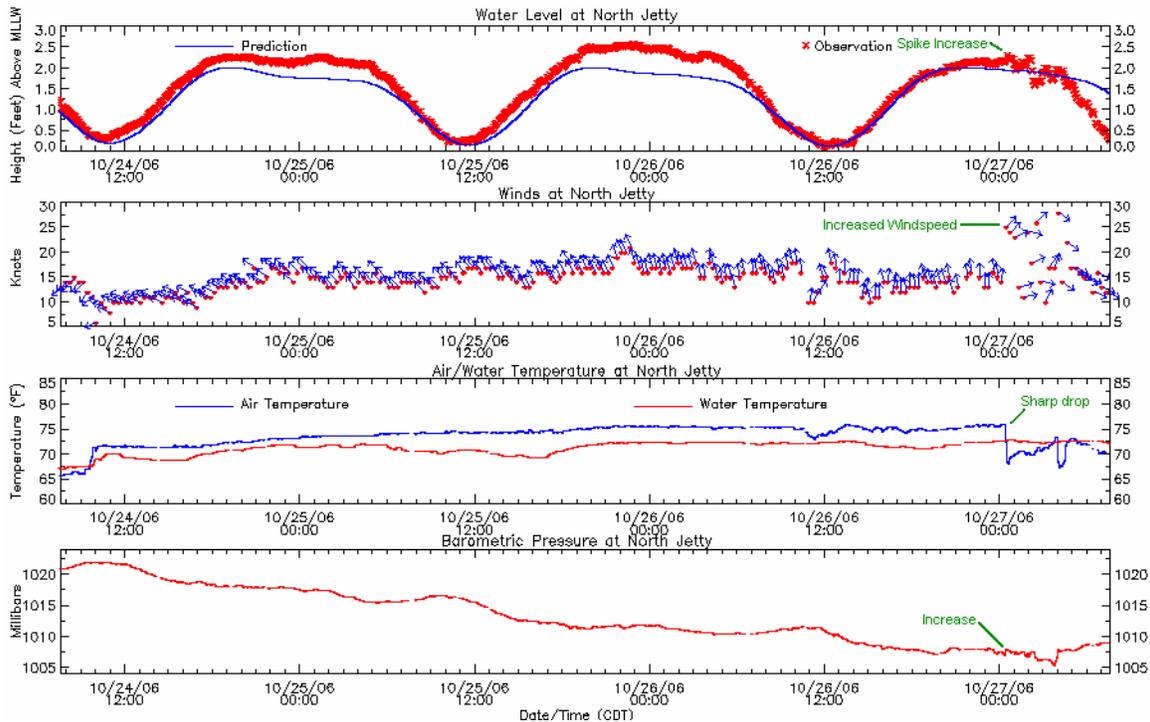

**Fig. 7** – 3-day meteorology from the PORTS station at North Jetty, TX, from 10/24 to 10/27. Significant changes are identified in green.

The WMSI values were collected at 5:00 GMT. The closest value to the location of the downburst was 84 (Fig. 8). The downburst occurred as a bow echo passed over the station towards the Gulf. A radar image of the area at 5:36 GMT verifies the convective activity (Fig. 9).

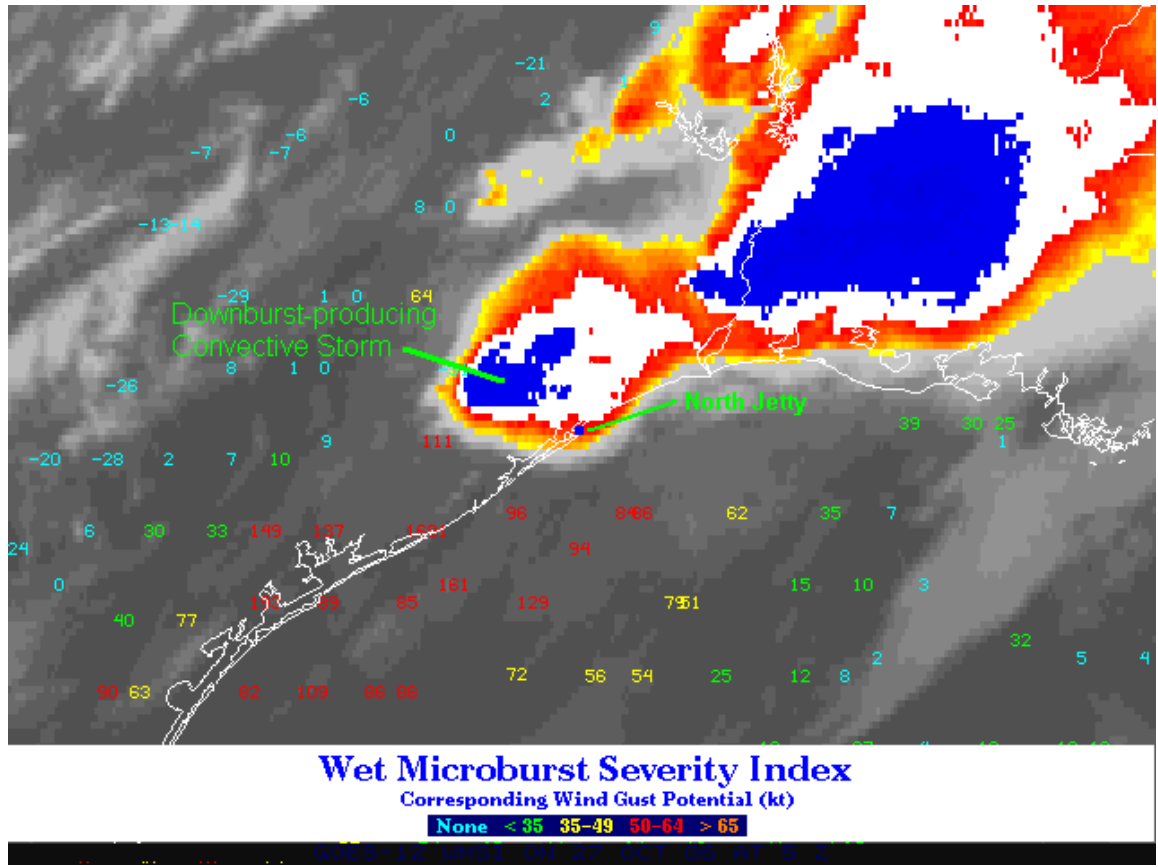

**Fig. 8** – WMSI values over the west Gulf coast at 5:00 GMT. The North Jetty station is located at the blue square. This image has been enhanced to show cloud top temperatures. The blue regions of clouds are between roughly -80 and -100 degrees Celsius, and it can be inferred that these reach the greatest altitude.

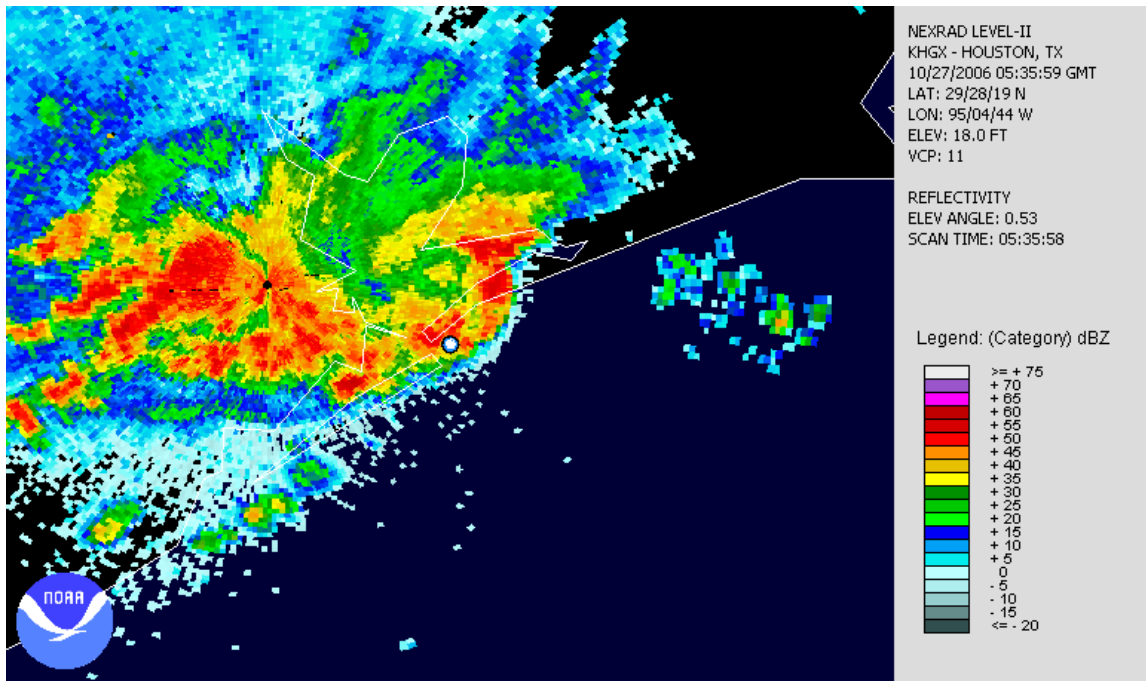

**Fig. 9** – Radar imagery from the NEXRAD station in Houston at 5:36 GMT. The North Jetty station is located at the white dot.

**Discussion**

The physical characteristics highlighted in this paper serve two purposes. The first is verification of microburst events. By looking at changes in factors of wind gust, wind direction, air temperature, and pressure, it is possible to determine the occurrence of a microburst. However, it is important to note that microbursts have a short duration, and therefore if these factors are sustained over long periods of time (more than thirty minutes) it is more likely an indication of storm activity.

A relatively high wind gust is one of the first signs observed in data recordings to identify a microburst event. The downbursts examined in this paper all reached wind gusts of thirty-five knots or more, although microbursts have been recorded at wind gusts little over 20 knots. (Geerts, 2000) A change in wind direction is also observed during a downburst event. The downbursts at Tolchester on September 28, 2006 and at North Jetty on October 13, 2006 have very large wind direction changes, both over one-hundred degrees.

Rapid air temperature decreases are another sign of downburst activity. The best representation of this is in the event at North Jetty on October 27, 2006. It appears that there may have actually been two events based on the wind patterns and the appearance of two sharp drops in temperature. The events at North Jetty studied in this paper both have temperature drops of roughly eight degrees in a only six to twelve minutes and the observed temperature at Tolchester dropped twelve degrees in the same amount of time. The drops in both cases are significant and can be attributed to a downburst event.

Pressure is also observed in each event rapidly increasing. It is expected that the pressure will rise between two millibars and five millibars. (Carcena, "Microbursts-…") Tolchester event reaches this expected interval. However, the events at North Jetty exhibit relatively low pressure increases. It is possible that the monitoring stations were not close enough to the downburst to record the complete pressure difference produced by the downburst. Despite the small size of the increase observed for these two events, the pattern of a pike surrounded by two troughs is present, which supports the idea that distance prevented the station from observing the full magnitude of the difference.

The other purpose of the observed physical characteristics is to serve as a comparison to the resulting marine conditions, which in this study are the changes in observed water level. In each case the maximum wind gust, the increase in water level, and the WMSI value are analyzed. Because of geographical difference, the Tolchester case and the two North Jetty cases are separated in some aspects of the analysis.

The Tolchester WMSI value is considerably lower than expected. A 49 knot wind is predicted theoretically by WMSI values of roughly 79. The recorded value of 45, however, predicts wind speeds of less than 35 knots based on the existing statistical relationship. This discrepancy can possibly be explained by two factors. The first is the geographic factor mentioned previously. When validating the WMSI, Pryor and Ellrod (2004) noted that the threshold used by the WMSI would require tuning between geographic regions. The other factor is that the GOES data is not available in thicker cloud cover, which is generally associated with convective activity. The closest WMSI value recorded may not be an accurate measurement of the event.

The two events at North Jetty have expected WMSI values. Although, like the Tolchester event, the WMSI values are not observed directly over the recording stations, it is clear in these images that the TeD and CAPE are at a significant level for wet microburst activity.

For the Tolchester case this difference is quite significant, roughly 0.84 feet. The two events at North Jetty are less, but also had very different wind speed conditions. The event on October 13, 2006 had only a max wind gust of 35 knots, which accounts for its 0.31 increase in water level. The October 27 event, however, only has about a 0.17 increase despite a 41 knot wind gust. This is most likely due to the elevated winds observed prior to the downburst event. The Tolchester and October 13 events both followed sustained winds of roughly 10 knots with gusts only a few knots higher. The October 27 event followed winds of 15 knots or more with 20 knot gusts, meaning that wave activity was already increased and resulted in less of an observed effect from the microburst winds. Another factor that illuminates this result is the irregular water level observations occurring after the downburst event.  The water level oscillates a great deal and it is possible that although the microburst winds did not cause a significant increase in water level, they may have sustained a much higher water level than that which would have otherwise occurred.  The water level decreases by 0.33 feet shortly after the downburst occurred, suggesting that in the absence of the downburst the water level would have already fallen.

**Conclusion**

There are several possible results of wet microburst activity in marine environments discussed in this paper.  It is clear that the short duration of a microburst allows for significant increases in water level.  With this information comes the possibility of significant increases in wave activity as well, which is a key hazard to marine craft, especially in port areas. Wind speeds of 35 to 40 knots are shown to be capable of water level differences of four inches over a 10 to 15 minute period.  Wind speeds of 45 to 50 knots over the same time period are capable of much larger increases, observed in this paper between 0.8 and 0.9 feet.  WMSI values that predict wind gusts of 35 to 50 knots are between 50 and 80.  (Pryor and Ellrod, 2004)  When observing values within this interval or above it, at least a four inch increase in water level can be expected in the event of a convective downburst.  Wind speeds and water level changes resulting from convective downbursts are shown to produce noticeable and potentially hazardous conditions for marine craft.